\documentclass{article}

\usepackage[nonatbib, final]{neurips_2022_ml4ps}

\usepackage[utf8]{inputenc} 
\usepackage[T1]{fontenc}    
\usepackage{hyperref}       
\usepackage{url}            
\usepackage{booktabs}       
\usepackage{amsfonts}       
\usepackage{nicefrac}       
\usepackage{microtype}      
\usepackage{xcolor}         

\usepackage{microtype}
\usepackage{graphicx}
\usepackage{subfigure}

\usepackage{amssymb}
\usepackage{amsmath}

\usepackage{wrapfig}

\usepackage[backend=biber,natbib=true, sorting=none]{biblatex}

\graphicspath{{figures/}}

\title{Discovering Long-period Exoplanets using Deep Learning with Citizen Science Labels}

\author{%
  Shreshth A. Malik\thanks{Correspondance to \texttt{\href{mailto:shreshth@robots.ox.ac.uk}{shreshth@robots.ox.ac.uk}}}
    \\
  OATML\\
  University of Oxford\\
   \And
  Nora L. Eisner
    \\
  Department of Physics\\
  University of Oxford\\
   \And
  Chris J. Lintott
    \\
  Department of Physics\\
  University of Oxford\\
   \And
  Yarin Gal
    \\
  OATML\\
  University of Oxford\\
}

\begin{document}

\maketitle

\begin{abstract}
Automated planetary transit detection has become vital to prioritize candidates for expert analysis given the scale of modern telescopic surveys.
While current methods for short-period exoplanet detection work effectively due to periodicity in the light curves, there lacks a robust approach for detecting single-transit events. 
However, volunteer-labelled transits recently collected by the Planet Hunters TESS (PHT) project now provide an unprecedented opportunity to investigate a data-driven approach to long-period exoplanet detection.
In this work, we train a 1-D convolutional neural network to classify planetary transits using PHT volunteer scores as training data. 
We find using volunteer scores significantly improves performance over synthetic data, and enables the recovery of known planets at a precision and rate matching that of the volunteers. Importantly, the model also recovers transits found by volunteers but missed by current automated methods.

\end{abstract}

\section{Introduction}
\label{sec:intro}
Astronomical datasets from recent survey missions such as TESS \cite{ricker2014transiting} are now large enough to make inspection by experts implausible. Discovering planets by observing the effect of their passage on their parent stars---the transit method---for example, requires automated analysis. However, incumbent algorithms that flag potential exoplanets 
rely on detecting periodic signals in the brightness of the star---its light curve (LC). This skews the observed distribution to short-period planets, resulting in fewer longer-period planets than expected in our catalog.
In response, citizen science projects have proven successful in finding planets that have been missed by automated detection methods \cite{fischer2012planet, eisner2021planet}. In Planet Hunters TESS\footnote{\url{http://www.planethunters.org}} (PHT), volunteers from the general public inspect LCs for planetary transits by eye. The data flagged by the volunteers has enabled a number of novel astronomical discoveries, notably of longer-period exoplanets \cite{eisner2021planet, eisner2022planet}. 

In this work, we investigate using volunteer transit confidence scores from PHT as soft labels to train a 1-D convolutional neural network (CNN) to detect single-transit events from TESS LCs. We find that training with volunteer scores as the main training signal enables better recovery of known planets compared to synthetic data, with a precision and recall similar to that of the volunteers. Moreover, the model is able to detect planets missed by traditional automated algorithms and even some that are missed by volunteers. 
The model could therefore serve as a basis for a human-in-the-loop machine learning pipeline for longer-period exoplanet discovery.




%

\section{Background and Related Work}

We focus on the \textit{detection} part of the exoplanet discovery pipeline, which involves identifying and short-listing high-potential exoplanet candidates from a large observational dataset for further analysis. 
We investigate detecting exoplanets using the transit method, where temporary decreases in brightness of stars can be used to identify a transiting planet.


\paragraph{Light Curves}

The observed brightness of a star varies due to both astrophysical and instrumental noise processes. These variations can include both periodic (e.g. stellar variability) and irregular signals (e.g. flares) which can either be confused for transits, or can confound automatic searches.
Contamination from other sources, the inherent variability of the target stars and telescope systematics can also cause anomalies and missing data. This means that classical signal processing struggles to robustly identify transits across the wide distribution of observed LCs. Figure \ref{fig:qual-eval} in Appendix \ref{sec:app-qual-eval} shows some examples of typical LCs. 


\paragraph{Traditional Exoplanet Detection and Validation}

The most common approach to automated detection is to search for periodic signals and use a box-fitting algorithm to find transit-like dips in the phase-folded LC \cite{kovacs2002box}. The flagged LCs are then validated using a combination of diagnostics, human vetting, probabilistic and machine learning approaches \cite{montet2015stellar, thompson2015machine, zucker2018shallow, shallue2018identifying,ansdell2018scientific, yu2019identifying, osborn2020rapid, armstrong2021exoplanet, valizadegan2022exominer}. 



\paragraph{Deep Learning for Exoplanet Detection}

Recent work has sought to go beyond validation and instead use deep learning directly on the (non-phase-folded) LCs to discover new candidates missed by detection algorithms. 
However, the comparatively small number of positive examples has limited the applicability of deep learning for exoplanet discovery  \cite{hinners2018machine}. Prior work trains on simulated data to overcome this bottleneck, but this has been shown to have limited applicability on real LCs due to the complex noise processes involved \cite{zucker2018shallow, malik2022exoplanet}.
\citet{olmschenk2021identifying} used a combination of real and synthetic data to construct a pipeline from full-frame images rather than the LCs. Nonetheless, the authors note that they still find a similar distribution of planets to those found by automated algorithms because the training data is biased towards multi-transit events. \citet{cui2021identify} present a promising approach of using 2-D object detection algorithms on images of the LC, but are again limited in training data by transits found by automated algorithms. In our work, by leveraging volunteer scores, we are able to recover planets outside of the distribution found by automated algorithms. 

\begin{figure}
    \centering
    \includegraphics[width=0.9\linewidth]{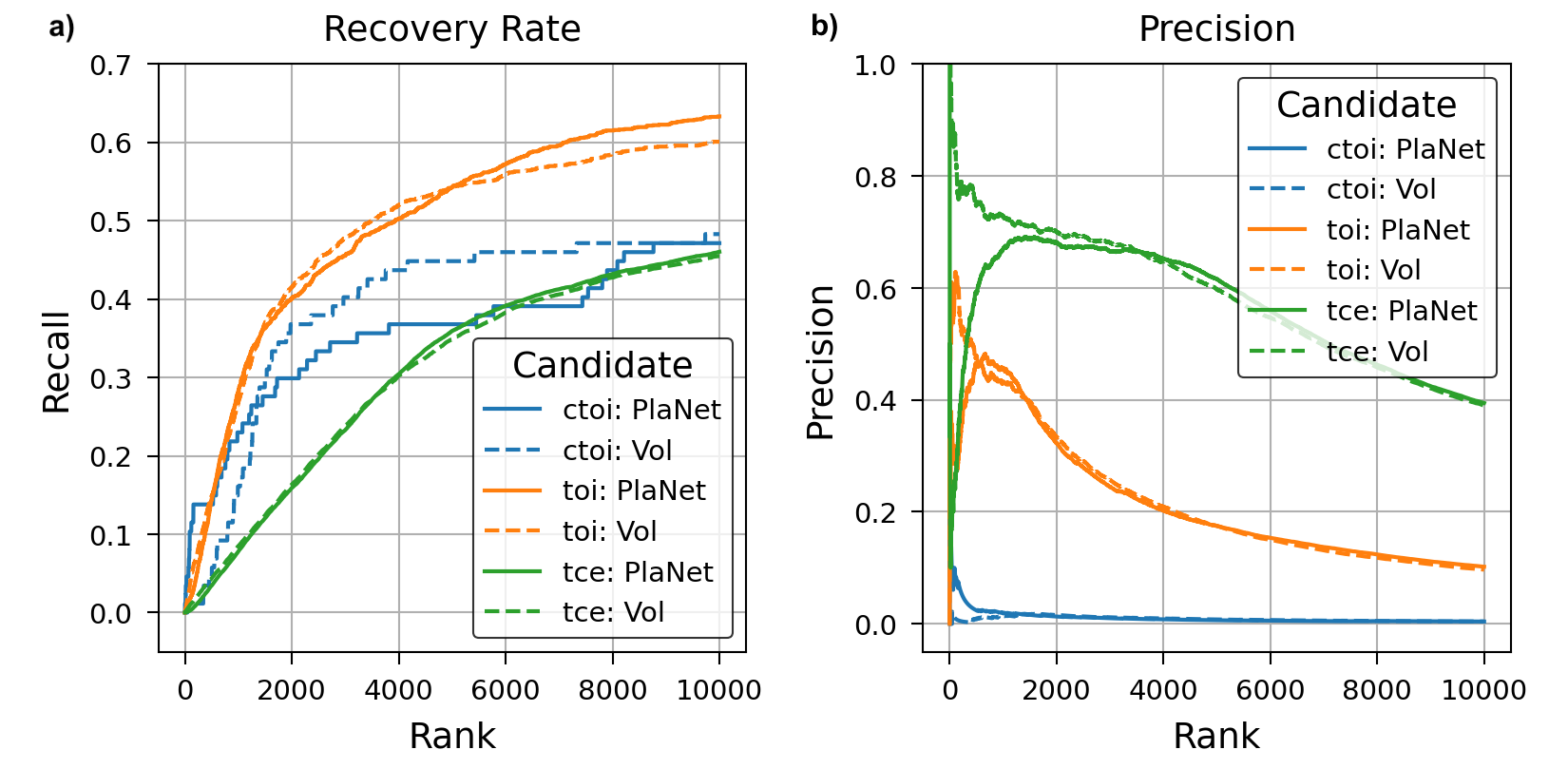}
    \caption{Known candidate recovery for the PlaNet model and volunteers on test sectors. \textbf{a)} Fractional recovery rate when test samples are ranked in order of predicted confidence. \textbf{b)} Precision of the top $k$ ranked predictions.}
    \label{fig:rec-rate}
\end{figure}

\section{Data}
\label{sec:data}


\subsection{TESS Light Curves}
\label{sec:lcs}

The TESS mission observes around 20,000 stars from a new sector of the sky every month. We used the two-minute cadence LCs from the SPOC pipeline \cite{jenkins2016tess}\footnote{\url{https://archive.stsci.edu/missions-and-data/tess}}.
To provide more realistic evaluation, we divided the train and test data by sector as the model will be used for prioritizing future observations rather than post-hoc analysis. Sectors 10-35 were used for training and validation, and sectors 36-43 for evaluation. Appendix \ref{sec:app-data} contains more details on the data.


\begin{table}[t]
\centering
\caption{Recovery of TOIs comparing volunteers to variations of PlaNet. The subscript denotes the proportion of synthetic planets and EBs used (default, 0.1). The recall (R) and precision (P) for the top $k$ ranked predictions, and the receiver operating characteristic and precision-recall area under curve (AUC, PR-AUC) metrics are given for each model.}
\begin{tabular}{@{}l|llll|llll|ll@{}}
\toprule
\textbf{Model}                                   & \multicolumn{10}{c}{\textbf{TOIs}}                                                             \\ \midrule \midrule
                            & R@100 & 1,000 & 5,000 & 10,000 & P@100 & 1,000 & 5,000 & 10,000 & AUC & PR-AUC \\ \midrule
Volunteers                                 & \textbf{0.030} & 0.268   & \textbf{0.543}   & 0.601    & \textbf{0.490} & 0.431   & \textbf{0.175}   & 0.097    & 0.799   & \textbf{0.261}  \\
\textbf{PlaNet}                                    & 0.019 & \textbf{0.283}   & \textbf{0.542}   & \textbf{0.633}    & 0.300 & \textbf{0.455}   & \textbf{0.175}   & \textbf{0.102}    & \textbf{0.835}   & 0.227  \\
PlaNet\textsubscript{0.0} & 0.023 & 0.241   & 0.525   & 0.594    & 0.370 & 0.388   & 0.169   & 0.096    & 0.810   & 0.197  \\
PlaNet\textsubscript{0.3} & 0.009 & 0.158   & 0.518   & 0.620    & 0.140 & 0.255   & 0.167   & 0.100    & 0.637   & 0.147  \\
PlaNet\textsubscript{all} & 0.009 & 0.113   & 0.396   & 0.517    & 0.140 & 0.182   & 0.127   & 0.083    & 0.756   & 0.091  \\ \bottomrule
\end{tabular}
\label{tab:main-results}
\end{table}

\paragraph{Pre-processing}
\label{sec:preprocess}
The LCs were pre-processed in three steps. \textbf{1) }\textbf{Anomaly removal}. We used the PDCSAP fluxes, which are nominally corrected for instrument variations and flux contamination from nearby stars. We also filtered using the \texttt{QUALITY} marker given in LC files to remove anomalies. \hfill \textbf{ 2)} \textbf{Binning}. 
For computational reasons we binned the data to a 14-minute cadence. This is the same binning factor for LCs shown to the volunteers. The duration of transits of interest are generally on the order of hours to days, so the characteristic shape of the dip is still identifiable at this resolution. Empirical tests confirmed that there was no significant performance difference in using LCs binned to 6-minutes or 14-minutes. \textbf{3) }\textbf{Normalisation}. We divided and subtracted by the median such that the LC is centred at zero. Dividing by the median (rather than the standard deviation) is common practice in LC analysis as this allows a comparison of the magnitude of brightness dips, which can differentiate between false positives and planets.
We truncated from the start and end (in random proportions) such that all LCs have a binned length of 2,500, and imputed missing values with zeros.

\paragraph{Synthetic Data}
\label{sec:synth-gen}

In this work we investigate the efficacy of using volunteer scores as a training signal for single-transit detection. We compare model performance when trained with varying amounts of additional synthetic data as a comparison. The majority of transit-like signals in LCs correspond to false positives ($>95\%$) \cite{sullivan2015transiting}. Thus to help differentiate these cases, we included
a proportion of synthetic data from the test ETE-6 dataset \cite{jenkins2018simulated} corresponding to transits and eclipsing binaries (EBs, the most common false positive) in equal proportions.
As we focus on single-transits, we only take one transit from each synthetic LC to inject into a random section of the base LC. We used the full flux for EBs as asymmetric dips are often used to identify them.

\subsection{Volunteer Scores and Planetary Candidate Labels}
\label{sec:pht-labels}

We used the aggregated confidence scores from PHT that take into account the volunteer skill level \cite{eisner2021planet} as soft targets to train the network. We also cross-referenced the star corresponding to each LC with discrete classifications of planetary candidates from TESS for evaluation. The TESS automated pipeline flags $\sim7\%$ of the observations as \textit{threshold crossing events} (TCEs) which indicates the presence of a periodic signal. TCEs are then analysed by the TESS team and $\sim18\%$ are promoted to \textit{TESS Objects of Interest} (TOIs) status as likely planetary candidates. Similarly, candidates that have been identified through volunteer flagging and vetted by the PHT science team are called \textit{Community} TOIs (cTOIs). Overall, $\sim$1\% of all LCs have planetary transits, whereas $\sim$21\% have a non-zero volunteer confidence scores. Thus the soft labels provide a better training signal to the model.


\section{Method}
\label{sec:method}

\paragraph{PlaNet Model}

A 1-D CNN we call \textit{PlaNet} was trained using a cross-entropy loss for binary classification of light curves with or without a planetary transit. The model is primarily based on the work by \citet{olmschenk2021identifying}.
In comparison to \citet{olmschenk2021identifying}, we used a deeper network (22 convolutional layers) and larger kernel sizes in the first two layers (7 and 5 respectively) without down-sampling. We also used residual connections to mitigate the vanishing gradient effect that is associated with deeper models  \cite{HeZRS16}. These modifications resulted in significant performance gains over the original model. Appendix \ref{sec:app-model} contains further details on the model. Our code is publicly available\footnote{\url{https://github.com/s-a-malik/pht-ml}}.

\paragraph{Data Augmentations}

During training, we added transformations to help with generalisation, each with a probability of 0.1. To simulate a noise process, we randomly chose another LC with volunteer score of 0.0 
to inject into the base LC. Three types of data shifts were also incorporated. 1) Two randomly chosen non-overlapping sections (each 25\% of the LC) were switched. 2) The LC was reversed temporally. 3) A random section (10\%) of the LC was deleted.

\section{Results and Discussion}
\label{sec:results}

\begin{figure}
    \centering
    \includegraphics[width=0.9\linewidth]{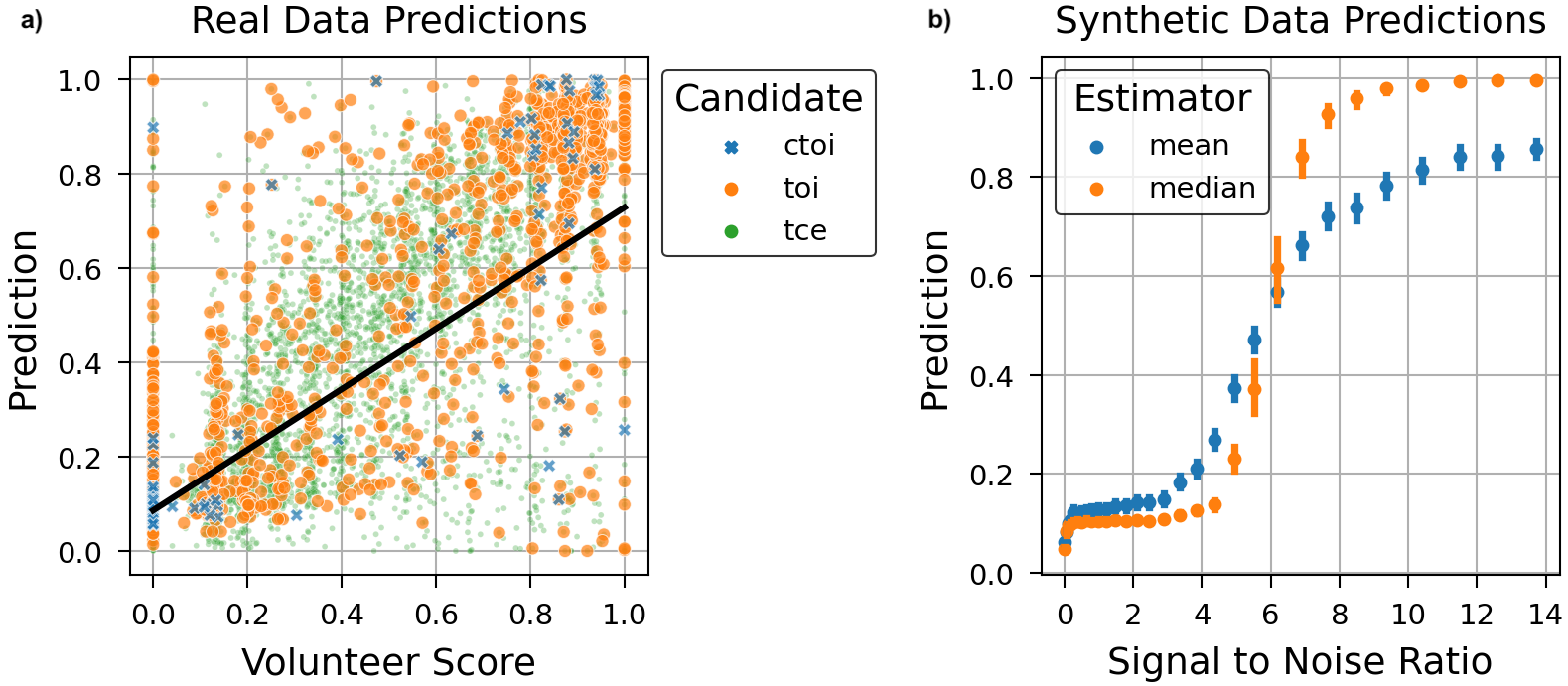}
    \caption{Predictions on real and synthetic test data. \textbf{a)} Parity plot for PlaNet predictions on real data. Null label predictions from volunteer scores are not plotted for clarity but included in the regression computation (black line). \textbf{b)} Predictions on injected transits as a function of signal to noise ratio. 
    A 95\% confidence interval is shown for each binned value.
    }
    \label{fig:scatter-synth}
\end{figure}

\paragraph{Using volunteer scores as training data enables effective and diverse planet recovery.} Figure \ref{fig:rec-rate} shows the fractional recovery rate and precision of planetary candidates for the top $\sim 10\%$ ranked predictions of PlaNet and volunteers. We find that PlaNet's performance on TOIs is similar to that of volunteers. Moreover, we find that the model recovers around a third of the cTOIs in the top 2\% ranked results, indicating that the model also finds planets that the TOI pipeline misses. 
Evaluation on synthetic data (Figure \ref{fig:scatter-synth}b) indicates that performance greatly increases for transits with a signal to noise ratio greater than 6. This is similar to that observed for volunteers \cite{eisner2021planet}.



\paragraph{Synthetic data can be useful but is not enough.} Table \ref{tab:main-results} shows TOI recovery performance of PlaNet when trained with different proportions of synthetic data. We find that training with volunteer scores significantly increases performance over training with only synthetic data. However the best performing model is one that does leverage a small amount of synthetic data (10\% synthetic transits, 10\% synthetic EBs, 80\% volunteer soft labels). We argue that the volunteer labels provides a strong training signal, but the additional hard labels provided by the synthetic data help identify transits and rule out false positives with more confidence. This leads to improved TOI recovery.

\paragraph{Comparing PlaNet and volunteer predictions.}
In general, we find that the TOIs and cTOIs have high model predictions (Figure \ref{fig:scatter-synth}a)\footnote{Note that the majority of (c)TOIs with both low volunteer and low model scores do not have transits in that particular LC. Instead these are identified when the same star is observed in other sectors. 
The results in Table \ref{tab:main-results} are thus an underestimate of true performance.
}. In some cases however, there is disagreement between volunteer and model predictions. A review of a selection of these disputed LCs by an expert author (NE) suggested that several ($\sim30$\%) of those identified by PlaNet but missed by volunteers were worth further investigation, suggesting that the network is providing suitable candidates for review and observational follow-up. We also found that fewer multi-transit planets were recovered, likely as we explicitly focus training on single-transit events. Appendix \ref{sec:app-qual-eval} contains further qualitative analysis.

\section{Conclusion}
\label{sec:conc}

In this work, we used volunteer scores from Planet Hunters TESS to train a 1-D CNN to detect planetary transits from TESS light curves. 
The model was found to match the original volunteer's recovery rate. Moreover, as observed in \citet{eisner2021planet}, we recover single-transit candidates that are missed by traditional pipelines, and even some that are missed by volunteers. The model could therefore serve as a basis for a human-in-the-loop pipeline for longer-period exoplanet discovery. In future work, we hope to use the model predictions alongside volunteer scores on new sectors to prioritize analysis. Incorporating calibrated uncertainty estimation would help identify costly overconfident predictions. This will also enable an active learning pipeline to be developed which can make better use of volunteer time \cite{settles2009active, walmsley2020galaxy}. 


\section*{Broader Impact}
\label{sec:broader-impact}

Citizen science enables more accessible participation in the scientific process. We believe involving the public in scientific endeavours is beneficial for the community, widens perspectives and accelerates progress. We see our work as a step towards human-in-the-loop machine learning pipeline where volunteer and expert time for tedious labelling and analysis tasks is reduced.
However, we must also note that systems trained on crowd-sourced data can amplify existing biases if they exist in the data. If there are systematic issues with volunteer labelling these can propagate to the model and potentially be detrimental to performance.

\section*{Acknowledgments and Disclosure of Funding}
SM acknowledges funding from EPSRC Centre for Doctoral Training in Autonomous Intelligent Machines and Systems (Grant No: EP/S024050/1). YG acknowledges funding from the Turing Fellowship (Grant No. EP/V030302/1). 
Some of the data presented in this paper was obtained from the Mikulski Archive for Space Telescopes (MAST).
Planet Hunters TESS is supported in part by the Alfred P. Sloan Foundation.

\printbibliography

\section*{Checklist}

\begin{enumerate}

\item For all authors...
\begin{enumerate}
  \item Do the main claims made in the abstract and introduction accurately reflect the paper's contributions and scope?
    \answerYes{}
  \item Did you describe the limitations of your work?
    \answerYes{See Section \ref{sec:results} and Appendix \ref{sec:app-qual-eval}.}
  \item Did you discuss any potential negative societal impacts of your work?
    \answerYes{See Broader Impact section}
  \item Have you read the ethics review guidelines and ensured that your paper conforms to them?
    \answerYes{}
\end{enumerate}

\item If you are including theoretical results...
\begin{enumerate}
  \item Did you state the full set of assumptions of all theoretical results?
    \answerNA{}
	\item Did you include complete proofs of all theoretical results?
    \answerNA{}
\end{enumerate}

\item If you ran experiments...
\begin{enumerate}
  \item Did you include the code, data, and instructions needed to reproduce the main experimental results (either in the supplemental material or as a URL)?
    \answerYes{See \url{https://github.com/s-a-malik/pht-ml}}
  \item Did you specify all the training details (e.g., data splits, hyperparameters, how they were chosen)?
    \answerYes{See Section \ref{sec:method} and Appendix \ref{sec:imp-dets}}
	\item Did you report error bars (e.g., with respect to the random seed after running experiments multiple times)?
    \answerNo{Uncertainty quantification will be done in further work (Section \ref{sec:conc})}
	\item Did you include the total amount of compute and the type of resources used (e.g., type of GPUs, internal cluster, or cloud provider)?
    \answerYes{See Appendix \ref{sec:app-model}}
\end{enumerate}

\item If you are using existing assets (e.g., code, data, models) or curating/releasing new assets...
\begin{enumerate}
  \item If your work uses existing assets, did you cite the creators?
    \answerYes{See Section \ref{sec:data} and Section \ref{sec:method}}
  \item Did you mention the license of the assets?
    \answerNA{}
  \item Did you include any new assets either in the supplemental material or as a URL?
    \answerNA{}
  \item Did you discuss whether and how consent was obtained from people whose data you're using/curating?
    \answerNA{}
  \item Did you discuss whether the data you are using/curating contains personally identifiable information or offensive content?
    \answerNA{}
\end{enumerate}

\item If you used crowdsourcing or conducted research with human subjects...
\begin{enumerate}
  \item Did you include the full text of instructions given to participants and screenshots, if applicable?
    \answerNA{Data was crowdsourced by the Planet Hunters TESS project. See \citet{eisner2021planet} for details.}
  \item Did you describe any potential participant risks, with links to Institutional Review Board (IRB) approvals, if applicable?
    \answerNA{}
  \item Did you include the estimated hourly wage paid to participants and the total amount spent on participant compensation?
    \answerNA{}
\end{enumerate}

\end{enumerate}


\appendix

\section{Qualitative Analysis}
\label{sec:app-qual-eval}

\begin{figure}[h]
\centering
    \includegraphics[width=\linewidth]{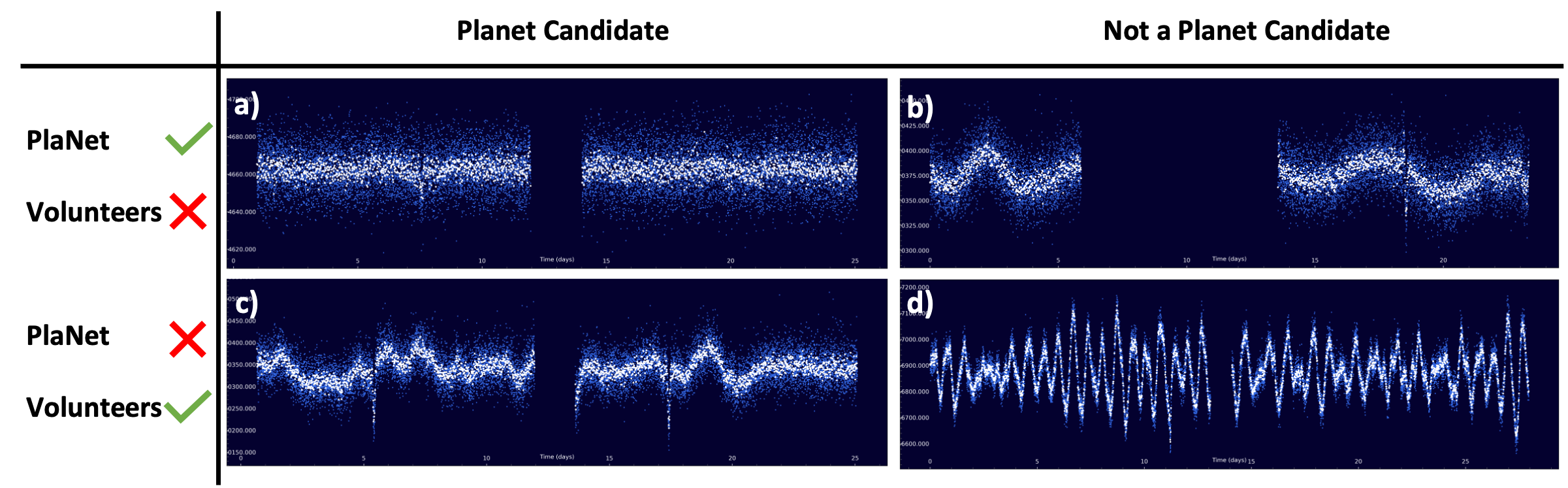}
    \caption{Example success and failure modes of volunteers and the PlaNet model. In particular, we look at examples of both false positive and false negative prediction of known planets, where PlaNet and the volunteers' predictions are the most different. Each panel shows binned (white) and un-binned (blue) TESS light curve data. 
    \textbf{a)} An example of a PlaNet true positive and volunteer false negative. Here PlaNet flagged a new potential candidate with a shallow dip that volunteers missed. \textit{TIC ID: 371596256, Sector: 36}.
    \textbf{b)} An example of a PlaNet true negative and volunteer false positive. The dip observed is due to a background event. \textit{TIC ID: 422506318, Sector: 42}.
    \textbf{c)} An example of a PlaNet false negative and volunteer true positive. \textit{TIC ID: 460140752, Sector: 36}.
    \textbf{d)} An example of a PlaNet false positive and volunteer true negative. 
    \textit{TIC ID: 295317859, Sector: 39}.
    }
    \label{fig:qual-eval}
\end{figure}

We can gain an insight into where PlaNet does well and where it fails by looking at example predictions on the test sectors. Concretely, we investigated cases where there are the largest discrepancies between model predictions and volunteer scores (i.e. the top left and bottom right corners of Figure \ref{fig:scatter-synth}a). Figure \ref{fig:qual-eval} shows examples of these cases.

\paragraph{PlaNet Success Modes}

Figure \ref{fig:qual-eval}a shows a case where volunteers rejected a promising candidate (likely due to the low signal to noise ratio), but PlaNet successfully flagged it as a likely transit. Figure \ref{fig:qual-eval}b shows a false positive from the volunteers. Here the dip is due to a background event, but may be mistaken for a transit to the untrained eye. This was correctly predicted to be an unlikely transit by the model.

\paragraph{PlaNet Failure Modes}

The bottom row of Figure \ref{fig:qual-eval} shows examples where volunteers correctly identify planetary candidates and false positives, whereas the model makes incorrect predictions with high confidence. We find the model sometimes rejects candidates with multi-transit events. As we have explicitly trained on single-transit synthetics, the model may mistake slight asymmetry in multi-transit dips as an eclipsing binary (Figure \ref{fig:qual-eval}c). We also find that false positives arise more often when there is a lot of stellar variability, as the model may mistake the sharp intrinsic variation for a transit (Figure \ref{fig:qual-eval}d).

\section{Implementation Details}
\label{sec:imp-dets}

\subsection{Data}
\label{sec:app-data}

The composition of each of the training, validation, and test data splits is given in Table \ref{tab:data-split}. Planets and EBs with TIC IDs that are multiples of 4 were chosen for synthetic data generation in the training set, and those with a remainder of 1 for the validation set. The rest were chosen for the test set. 

When synthetic data was used, an equal proportion of synthetic transits to synthetic EBs was used in all cases. In the synthetic data composition experiments (Section \ref{sec:results}), the model that used only synthetic data used 30\% synthetic transits, 30\% EBs, and 40\% LCs which volunteers scored 0.0 which indicates they are unlike to contain a transit. On synthetic injection, we randomly selected a transit or EB from the relevant data split and point-wise multiplied it to the unnormalized base LC. We used base LCs that have a volunteer score of 0.0 such that they are unlikely to already contain a transit, but these may contain EBs. We further ensured that at least 80\% of the section of the base LC where the transit is being injected into was not missing data. This is to prevent injecting a transit into a missing data region where it would not be visible and thus be mislabelled. LC noise was injected in a similar way using median-normalised injection curves which had a volunteer score of 0.0.

The signal to noise ratio (SNR) for synthetic data was calculated as the depth of the injected transit divided by the Combined Differential Photometric Precision
(CDPP) of the base light curve. CDPP is the metric that defines the ease with which these weak terrestrial transit signatures can be detected. This is given at a series of durations (0.5, 1, 2 days). The closest one to the duration of the transits was used to calculate the SNR. Unlike in PHT and other work \cite{eisner2021planet, cui2021identify}, we did not place a minimum SNR constraint on injected transits as we hope to be able to identify shallow transits. We did, however, place a maximum SNR constraint of 15, and a maximum duration of 4 days as in PHT.

\subsection{Models}
\label{sec:app-model}

\begin{figure}
\centering
    \includegraphics[width=\linewidth]{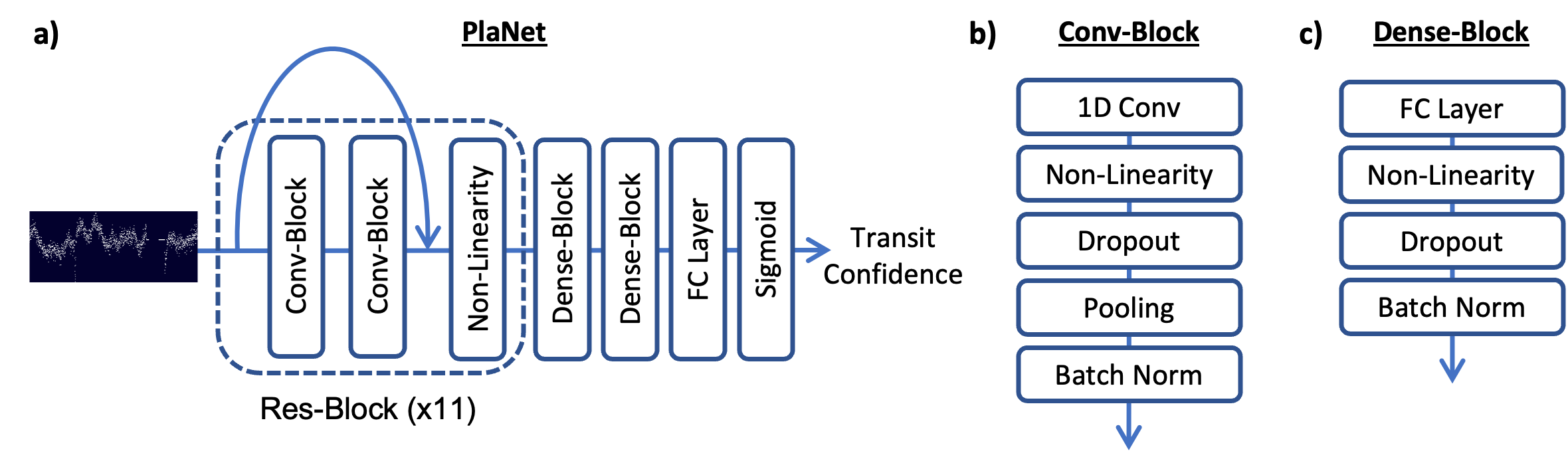}
    \caption{Simplified schematic of the proposed PlaNet architecture (inspired by \citet{olmschenk2021identifying}). \textbf{a)}  PlaNet features a series of convolutional blocks (Conv-Blocks) with residual connections \cite{HeZRS16}, followed by two Dense-Blocks. \textbf{b)} A Conv-Block. \textbf{c)} A Dense-Block.}
    \label{fig:model-diag}
\end{figure}

Figure \ref{fig:model-diag} shows a schematic of the proposed PlaNet architecture. PlaNet consists of 11 residual blocks, each consisting of two convolutional blocks with a skip connection \cite{HeZRS16}. These are followed by 2 dense blocks and finally a fully-connected layer with sigmoid activation to output a transit confidence score between 0 and 1. 
We do not use a fully convolutional model as some degree of temporal awareness is required to identify EB false positives with multiple dips. 

Conv-Blocks consists of a 1-D convolution, followed by a Leaky-ReLU non-linearity, dropout \cite{SrivastavaHKSS14}, a max-pooling operation and finally batch normalisation \cite{IoffeS15}. A generic Dense-Block consists of a linear layer, followed by a Leaky-ReLU, dropout and batch normalisation. Some operations in Conv and Dense-Blocks are not used depending on the layer of the network (Appendix \ref{sec:imp-dets}). In Res-Blocks, the second Conv-Block does not use a non-linearity. Instead the non-linearity is applied after the residual is added to its output. When pooling is used to downsample, a 1-D convolutional with kernel size of 1 and a stride equal to the pooling size is used to map the residual output to the same dimensions as the second Conv-Block output. 

Models were implemented in \textit{PyTorch} \cite{PaszkeGMLBCKLGA19}. The main PlaNet model hyperparameters are given in Table \ref{tab:hypers}. Architecture and optimization hyperparameters were chosen heuristically through minimising validation loss. Training runs on the full dataset took 15-24 hours on a single NVIDIA GeForce GTX 2080 Ti GPU. 

\begin{table*}[t]
\centering
\caption{Breakdown of the examples in each of the data splits used in this work. Strong volunteer scores are those that have an aggregated confidence over 0.5. The number of synthetic transits and eclipsing binaries used to generate the synthetic data is also given.}
\begin{tabular}{@{}lllllll@{}}
\toprule
\textbf{Data Split} & \textbf{Sectors} & \textbf{\begin{tabular}[c]{@{}l@{}}Total Light \\ Curves\end{tabular}}& \textbf{\begin{tabular}[c]{@{}l@{}}Non-zero \\ Vol. Scores\end{tabular}} & \textbf{\begin{tabular}[c]{@{}l@{}}Strong \\ Vol. Scores\end{tabular}} & \textbf{\begin{tabular}[c]{@{}l@{}}Synthetic\\ Transits\end{tabular}} & \textbf{\begin{tabular}[c]{@{}l@{}}Synthetic\\ EBs\end{tabular}} \\ \midrule
Training & 10--29 & 367,417 & 58,721 & 8,311 & 278 & 65 \\
Validation & 30--35 & 118,344 & 29,850 & 2,974 & 316 & 69 \\
Test & 36--43 & 137,657 & 44,891 & 4,662 & 616 & 145 \\ \bottomrule
\end{tabular}
\label{tab:data-split}
\end{table*}

\begin{table}[h]
\caption{Hyperparameter configuration for PlaNet.}
\begin{tabular}{@{}p{0.35\linewidth}p{0.64\linewidth}@{}}
\toprule
\textbf{Hyperparameter}                        & \textbf{Value}                                                                 \\ \midrule \midrule
\textbf{Optimization}                          &                                                                       \\ \midrule 
Optimizer                             & AdamW \cite{adamw}                                   \\
L2 Weight Regularisation              & 0.01                                                                  \\
Learning Rate                         & 0.001                                                                  \\
Batch Size                            & 256                                                                   \\
Early Stopping Patience (epochs)               & 300                                                                   \\
Dropout Rate                          & 0.1                                                                   \\
Max Epochs                            & 1000                                                                  \\ \midrule \midrule
\textbf{Architecture}                          &                                                                       \\ \midrule 
Hidden Layer Non-linearities          & Leaky-ReLU (slope=0.01)                                               \\
Layers                                & {[}Res-Block (x11), Dense-Block (x2), Linear{]}                       \\
Kernel Size                           & {[}7, 5, 3, 3, 3, 3, 3, 3, 3, 3, 3, N/A, N/A, N/A{]}                  \\
Dropout (True/False)                  & {[}F, T, T, T, T, T, T, T, T, T, T, T, F, N/A{]}                      \\
Batch Normalisation (True/False)      & {[}F, T, T, T, T, T, T, T, T, T, T, T, F, N/A{]}                      \\
Number of Out Channels/Units          & {[}32, 32, 32, 64, 64, 128, 128, 128, 128, 128, 128, 1280, 256, 20{]} \\
Max Pooling Size                      & {[}1, 1, 2, 2, 2, 2, 2, 2, 2, 2, 1, N/A, N/A, N/A{]}                  \\
Total Number of Parameters            & 1,054,889                                                               \\ \midrule \midrule
\textbf{Data}                                  &                                                                       \\ \midrule 
Synthetic Transit Proportion          & 0.1                                                                   \\
Synthetic EB Proportion & 0.1                                                                   \\
Training Augmentation Probability              & 0.1                                                                   \\
Bin Factor (Max LC Length)            & 7 (2500)                                                              \\
Permute Fraction in Training          & 0.25                                                                  \\
Delete Fraction in Training           & 0.1                                                                   \\ \bottomrule
\end{tabular}
\label{tab:hypers}
\end{table}

\end{document}